\documentclass[amssymb,twocolumn,eqsecnum,aps]{revtex4}
\usepackage{amsmath}
\begin{document}
\title{Quantum Theory of Electrodynamics in Linear Media Subject to
Boundary Conditions}
\author{Michael E. Crenshaw}
\affiliation{AMSRD-AMR-WS-ST, USA RDECOM, Aviation and Missile RDEC,
Redstone Arsenal, AL 35898, USA}
\date{\today}
\begin{abstract}
We show that the material dependencies of macroscopically quantized
fields in linear media are not consistent with the classical
electromagnetic boundary conditions.
We then phenomenologically construct macroscopic quantized fields that
satisfy quantum--classical correspondence with the result indicating
that the canonical momentum in a linear medium is modified as a
consequence of the reduced speed of light.
We re-derive D'Alembert's principle and Lagrange's equations
for an arbitrarily large region of space in which light signals travel
slower than in the vacuum and show that the resulting modifications to
Lagrangian dynamics, including the canonical momenta, repair the
violation of the correspondence principle for macroscopically quantized
fields.
\end{abstract}
\maketitle
\vskip 3.0cm
\par
\section{Introduction}
\par
Quantum electrodynamics describes the quantum mechanical properties of 
electromagnetic fields in the vacuum.
The analogous theory for the quantum optical properties of electric and
magnetic fields in a dielectric or similar linear medium can be obtained
by quantizing the macroscopic fields in the continuous
medium \cite{BIGinz,BIJauch,BIDrummond,BIMilonni,BIGlauber,BIMarcuse},
as well as by applying quantum electrodynamics to a microscopic model of
the medium \cite{BIHop,BIKnoesM,BIHB,BICMomSp}.
Adopting the former approach in 1940, Ginzburg \cite{BIGinz} quantized
the macroscopic electromagnetic field in a simple transparent linear
dielectric.
The idealized dielectric studied by Ginzburg is a model for a more
complex material that exhibits dispersion, absorption, nonlinearity,
and other extended effects to varying degrees.
The difficulties of treating dispersion within the macroscopic
quantization procedure have been addressed by
Jauch and Watson \cite{BIJauch}, Drummond \cite{BIDrummond}, and
Milonni \cite{BIMilonni}.
Huttner and Barnett \cite{BIHB} treated absorptive dielectrics using a
macroscopic quantization procedure based on damped Hopfield polaritons.
Quantization of the macroscopic field in nonlinear
dielectrics \cite{BIHillMlod},
magnetodielectric matter \cite{BImagnetodiel}, 
and negative-index materials \cite{BInegindex}
has also been investigated.
In contrast to reports of extended aspects of quantized fields in
matter, the current work challenges the validity of the macroscopic
quantization procedure itself.
Specifically, we use the classical electromagnetic boundary conditions
to show that the quantized field inside a linear medium, as it is 
currently derived
\cite{BIGinz,BIJauch,BIDrummond,BIMilonni,BIGlauber,BIMarcuse},
violates quantum--classical correspondence.
\par
Materials are of finite extent and we are usually interested in how the
dynamics of a field change with the properties of the medium.
In the limit of large numbers, the quantum and classical relations must
have the same form.
We find, however, that the material dependencies of macroscopically
quantized fields in linear media are not consistent with the classical
electromagnetic boundary conditions.
Subsequently, the boundary conditions are used to phenomenologically
construct quantized fields in matter that are expressed in
terms of material-independent creation and annihilation operators
so that quantized fields in different media are related by the overall
normalization that coincides with the classical boundary conditions.
The generalized position and momentum operators of the conforming
theory do not obey standard commutation relations.
We re-derive D'Alembert's principle and Lagrange's equations based on
the relativity of dynamics in a linear medium \cite{BIFinn,BIFinn1} and
show that the reduced speed of light changes the definition of the
canonically conjugate momentum.
Then we incorporate the modified conjugate momentum into the
quantization procedure
\cite{BIGinz,BIJauch,BIDrummond,BIMilonni,BIGlauber,BIMarcuse}
and derive quantized fields that are the same as the phenomenological
fields and likewise satisfy the correspondence principle.
\par
\section{Quantization Procedure}
\par
A typical Ginzburg macroscopic quantization procedure \cite{BIMarcuse}
is based on an expansion of the vector potential in terms of modes
\begin{equation}
{\bf A}=c\sum_{l\lambda} q_{l\lambda}(t){\bf u}_{l\lambda}({\bf r})
\label{EQu2.01}
\end{equation}
in a dielectric subject to the usual continuum assumption of an
arbitrarily large, linear, isotropic, homogeneous, transparent
medium with refractive index $n$.
All fields, variables, and operators are macroscopic in the sense of
representing quantities as continuum averages over a volume larger than 
a cubic wavelength.
The ${\bf u}_{l\lambda}$ are orthonormal functions that satisfy periodic
boundary conditions on the quantization volume.
Substituting Eq.\ (\ref{EQu2.01}) into the electromagnetic Lagrangian
\begin{equation}
L=\frac{1}{2}\int
\left (\frac{n^2}{c^2} \left ( \frac{d{{\bf A}}}{d t}
\right )^2-(\nabla\times {\bf A})^2 \right ) d^3{\bf r}
\label{EQu2.02}
\end{equation}
and the wave equation, applying separation of variables and the identity
\begin{equation}
\int_V (\nabla\times{\bf u}_{l\lambda}) \cdot
(\nabla\times{\bf u}_{l^{\prime}\lambda^{\prime}}) d^3{\bf r}
= \int_V {\bf u}_{l\lambda}\cdot
[\nabla\times(\nabla\times{\bf u}_{l^{\prime}\lambda^{\prime}})]
d^3{\bf r},
\label{EQu2.03}
\end{equation}
one obtains the Lagrangian
\begin{equation}
L=\frac{1}{2}
\sum_{l\lambda}
\left ( n^2\dot q_{l\lambda}^2
-n^2\omega_l^2 q_{l\lambda}^2\right )
\label{EQu2.04}
\end{equation}
using orthonormality of the mode functions.
The conjugate momenta \cite{BIMarcuse,BIGoldstein,BIMarion}
\begin{equation}
p_{l\lambda}=\frac{\partial L}{\partial\dot q_{l\lambda}}
=n^2\dot q_{l\lambda}
\label{EQu2.05}
\end{equation}
are then used to construct the effective Hamiltonian
\begin{equation}
H=\frac{1}{2}\sum_{l\lambda} \left ( \frac{{p}_{l\lambda}^2}{n^2}+
n^2\omega_l^2 q_{l\lambda}^2\right ).
\label{EQu2.06}
\end{equation}
The effective Hamiltonian takes the form
\begin{equation}
H=\frac{1}{2}\sum_{l\lambda} \left ( P_{l\lambda}^2+
\omega_l^2 Q_{l\lambda}^2\right )
\label{EQu2.07}
\end{equation}
by application of the canonical transformation
\begin{subequations}
\label{EQu2.08}
\begin{equation}
P_{l\lambda}=p_{l\lambda}/n
\label{EQu2.08a}
\end{equation}
\begin{equation}
Q_{l\lambda}=n q_{l\lambda}.
\label{EQu2.08b}
\end{equation}
\end{subequations}
\par
Quantization is accomplished by treating $P_{l\lambda}$ and
$Q_{l\lambda}$ as operators satisfying the commutation relations
\begin{equation}
[q_{l\lambda},p_{l^{\prime}\lambda}]
[Q_{l\lambda},P_{l^{\prime}\lambda}]
=i\hbar\delta_{ll^{\prime}\lambda\lambda^{\prime}}.
\label{EQu2.09}
\end{equation}
One then introduces the creation operators $\zeta^{\dagger}_{l\lambda}$
and the annihilation operators $\zeta_{l\lambda}$ by
\begin{subequations}
\label{EQu2.10}
\begin{equation}
P_{l\lambda}=i\sqrt{\frac{\hbar\omega_l}{2}}(\zeta^{\dagger}_{l\lambda}
-\zeta_{l\lambda})
\label{EQu2.10a}
\end{equation}
\begin{equation}
Q_{l\lambda}=\sqrt{\frac{\hbar}{2\omega_l}}
(\zeta_{l\lambda}^{\dagger}+\zeta_{l\lambda}),
\label{EQu2.10b}
\end{equation}
\end{subequations}
where the operators obey boson commutation relations
\begin{equation}
[\zeta_{l\lambda},\zeta_{l^{\prime}\lambda^{\prime}}]=
[\zeta_{l\lambda}^{\dagger},
\zeta_{l^{\prime}\lambda^{\prime}}^{\dagger}]=0
\; ; \;
[\zeta_{l\lambda},\zeta_{l^{\prime}\lambda^{\prime}}^{\dagger}]
=\delta_{ll^{\prime}}\delta_{\lambda\lambda^{\prime}}.
\label{EQu2.11}
\end{equation}
The effective, or macroscopic, Hamiltonian is therefore
\cite{BIGinz,BIJauch,BIDrummond,BIMilonni,BIGlauber,BIMarcuse,BIHop,
BIKnoesM,BIHB,BICMomSp}
\begin{equation}
H= \frac{1}{2}\sum_{l\lambda}\hbar\omega_l
\left ( \zeta_{l\lambda}^{\dagger} \zeta_{l\lambda}+
 \zeta_{l\lambda} \zeta_{l\lambda}^{\dagger}\right ) .
\label{EQu2.12}
\end{equation}
It is then straightforward to derive the
traveling-wave representation of the medium-assisted vector
potential operator
\cite{BIGinz,BIJauch,BIDrummond,BIMilonni,BIGlauber,BIMarcuse,BIHop,
BIKnoesM,BIHB,BICMomSp}
\begin{equation}
 {\bf A}=c
\sum_{l\lambda}
\sqrt{\frac{\hbar}{2n^2\omega_l V}}
\left ( \zeta_{l\lambda} e^{i{\bf k}_l\cdot{\bf r}} +
 \zeta_{l\lambda}^{\dagger} e^{-i{\bf k}_l\cdot{\bf r}} 
\right )
\hat {\bf e}_{{\bf k}_{l\lambda}}.
\label{EQu2.13}
\end{equation}
The effective Hamiltonian (\ref{EQu2.12}) and the macroscopic vector
potential operator (\ref{EQu2.13}) are the principle products of the
quantization procedures that appear in the macroscopic quantum
electrodynamics literature
\cite{BIGinz,BIJauch,BIDrummond,BIMilonni,BIGlauber,BIMarcuse,BIHop,
BIKnoesM,BIHB,BICMomSp}.
\par
The normalization of the expansion of the vector potential in
Eq.\ (\ref{EQu2.01}) is arbitrary.
It is sometimes customary to use a normalization of $1/n$ or
$1/\sqrt{\varepsilon}$ in Eq.\ (\ref{EQu2.01}).
This is equivalent to the canonical transformation (\ref{EQu2.08}) and
eliminates a step in the derivation.
No normalization was used here in order to facilitate comparison 
with a similar quantization procedure that is presented later.
\par
\section{Boundary Conditions}
\par
In the limit of large numbers, quantum and classical relations must have
the same form.
The nature of macroscopic fields as averages means that the limit will
be satisfied.
Classically, the spatial extent of a field inside a material is reduced
in width by a factor of $n$ compared to the width in vacuum due to the
reduced speed at which light travels in the medium.
The refracted field is then reduced in amplitude by $\sqrt{n}$ compared
to the incident field from vacuum as a requirement of conservation of
electromagnetic energy in the limit that reflection is negligible or
suppressed by an anti-reflection coating. 
\par
We consider an electromagnetic field normally incident on a
minimally reflective interface between the vacuum and a linear medium.
Forward propagating plane-wave solutions of the wave equation can be
represented by the vector potential
\begin{equation}
{{\bf A}}_f= A_f \cos{(-\omega t +k z +\phi)}\hat {\bf e}_i.
\label{EQm3.01}
\end{equation}
Here, $\hat {\bf e}_i$ is a unit vector transverse to the direction
of propagation and ${k=n\omega/c}$.
The amplitudes of the vector potential for the incident and
refracted fields are respectively denoted as $A_i$ and $A_t$.
\par
The Poynting--Umov vector $c{\bf E}\times{\bf B}$ is the continuous
energy flux vector that is associated with energy conservation.
In terms of vector potential amplitudes,
\begin{equation}
A_i^2=nA_t^2
\label{EQm3.02}
\end{equation}
is obtained from continuity of the Poynting--Umov vector in the
plane-wave cw limit.
We write
\begin{equation}
A_t=\frac{A_v}{\sqrt{n}}
\label{EQm3.03}
\end{equation}
for the relation between the refracted field $A_t$ and the field
$A_i=A_v$ that is incident from the vacuum.
Because fields can be transferred from one material to another through
index-matching or other anti-reflection techniques, 
\begin{equation}
\sqrt{n_1}{A_1}=\sqrt{n_2}{A_2}
\label{EQm3.04}
\end{equation}
is the relation between vector potential amplitudes in different
materials and is equivalent to the Fresnel relation in the limit
that the change in refractive index is sufficiently small that
reflections are negligible.
\par
We write the vector potential operator (\ref{EQu2.13}) as
\begin{equation}
 {\bf A}=c
\sum_{l\lambda}
\sqrt{\frac{\hbar}{2n\omega_l V}}
\left (\frac{\zeta_{l\lambda}}{\sqrt{n}}e^{i{\bf k}_l\cdot{\bf r}}+
\frac{\zeta_{l\lambda}^{\dagger}}{\sqrt{n}}
e^{-i{\bf k}_l\cdot{\bf r}} \right )
\hat {\bf e}_{{\bf k}_{l\lambda}} 
\label{EQm3.05}
\end{equation}
and define macroscopic annihilation and creation operators
\begin{subequations}
\label{EQm3.06}
\begin{equation}
a_{l\lambda}=\frac{ \zeta_{l\lambda}}{\sqrt{n}}
\label{EQm3.06a}
\end{equation}
\begin{equation}
a_{l\lambda}^{\dagger}
=\frac{ \zeta_{l\lambda}^{\dagger}}{\sqrt{n}}.
\label{EQm3.06b}
\end{equation}
\end{subequations}
By substitution from Eqs.\ (\ref{EQm3.06}), the macroscopic
vector potential operator (\ref{EQm3.05}) becomes
\begin{equation}
 {\bf A}=c
\sum_{l\lambda}
\sqrt{\frac{\hbar}{2n\omega_l V}}
\left ({ a_{l\lambda}}e^{i{\bf k}_l\cdot{\bf r}}+
{ a_{l\lambda}^{\dagger}}
e^{-i{\bf k}_l\cdot{\bf r}} \right )
\hat {\bf e}_{{\bf k}_{l\lambda}} .
\label{EQm3.07}
\end{equation}
In this form, the material dependence of the classical boundary
condition (\ref{EQm3.04}) is satisfied by the normalization
prefactor of the field
in which case the annihilation and creation operators
obey material-independent boson commutation relations
\begin{equation}
[ a_{l\lambda}, a_{l^{\prime}\lambda^{\prime}}]=
[ a_{l\lambda}^{\dagger},
 a_{l^{\prime}\lambda^{\prime}}^{\dagger}]=0 \; ; \;\;\;
[ a_{l\lambda}, a^{\dagger}_{l^{\prime}\lambda^{\prime}}]=
\delta_{l l^{\prime}} \delta_{\lambda \lambda^{\prime}},
\label{EQm3.08}
\end{equation}
while the polariton operators obey material-dependent commutation
relations
\begin{equation}
[ \zeta_{l\lambda}, \zeta_{l^{\prime}\lambda^{\prime}}]=
[ \zeta_{l\lambda}^{\dagger},
 \zeta_{l^{\prime}\lambda^{\prime}}^{\dagger}]=0 \; ; \;\;\;
[ \zeta_{l\lambda},
 \zeta^{\dagger}_{l^{\prime}\lambda^{\prime}}]=
n\delta_{l l^{\prime}} \delta_{\lambda \lambda^{\prime}}.
\label{EQm3.09}
\end{equation}
This result contradicts the condition (\ref{EQu2.11}) used in the
macroscopic Ginzburg quantization procedure in Section II.
Using the operators (\ref{EQm3.06}), one can show that the
commutator (\ref{EQu2.09}) should be
\begin{equation}
[Q_{l\lambda},P_{l^{\prime}\lambda^{\prime}}]
=[q_{l\lambda},p_{l^{\prime}\lambda^{\prime}}]
=in\hbar\delta_{ll^{\prime}\lambda\lambda^{\prime}},
\label{EQm3.10}
\end{equation}
based on the polariton commutation relations (\ref{EQm3.09}). 
The scaling of the commutator in Eq.\ (\ref{EQm3.10}) proves that the
generalized momentum and position variables are not canonically
conjugate in the usual sense.
\par
\section{Lagrange's Equations in Filled Spacetime}
\par
In order to derive the generalized momentum variables that are
canonically conjugate to the generalized coordinates in a linear medium,
we treat space as being entirely filled with an isotropic homogeneous
continuous medium that responds linearly to electromagnetic radiation. 
Starting from first principles, we then derive the characteristics of
relativity and Lagrangian dynamics for the case in which the effective
speed of light is $c/n$.
\par
Consider an inertial reference frame $S(t,x,y,z)$ with orthogonal axes
$x$, $y$, and $z$. Position vectors in $S$ are denoted by
${\bf x}=(x,y,z)$.
If a light pulse is emitted from the origin at time $t=0$, then
\begin{equation}
x^2+y^2+z^2-\left ( \frac{c}{n} t\right )^2=0 
\label{EQu4.01}
\end{equation}
describes wavefronts in the $S$ system.
Writing time as a spatial coordinate $ct/n$, the
four-vector \cite{BIFinn1}
\begin{equation}
{\mathbb X}=(ct/n,{\bf x})=(ct/n,x,y,z)
\label{EQu4.02}
\end{equation}
represents the position of a point in a dielectric-filled
four-dimensional spacetime.
\par
Now consider two inertial reference frames, $S(t,x,y,z)$ and
$S^{\prime}(t^{\prime},x^{\prime},y^{\prime},z^{\prime})$, in a
standard configuration \cite{BIRindler,BISchwarz,BICarroll}
in which $S^{\prime}$ translates at a constant
velocity $v$ in the direction of the positive $x$ axis and the origins
of the two systems coincide at time $t=t^{\prime}=0$.
If a light pulse is emitted from the common origin at time $t=0$, then
\begin{equation}
(x^{\prime})^2+ (y^{\prime})^2+ (z^{\prime})^2
-\left ( \frac{c}{n} t^{\prime} \right )^2=0 
\label{EQu4.03}
\end{equation}
describes wavefronts in the $S^{\prime}$ system and Eq.\ (\ref{EQu4.01})
holds for wavefronts in $S$.
The material Lorentz transformation \cite{BIFinn1}
\begin{subequations}
\label{EQu4.04}
\begin{equation}
x=\gamma(x^{\prime}+vt^{\prime})
\label{EQu4.04a}
\end{equation}
\begin{equation}
y=y^{\prime}
\label{EQu4.04b}
\end{equation}
\begin{equation}
z=z^{\prime}
\label{EQu4.04c}
\end{equation}
\begin{equation}
t=\gamma\left (t^{\prime}+\frac{n^2v}{c^2}x^{\prime}\right )
\label{EQu4.04d}
\end{equation}
\end{subequations}
is derived by the usual methods \cite{BIRindler,BISchwarz,BICarroll},
where
\begin{equation}
\gamma=\frac{1}{\sqrt{1-\frac{n^2v^2}{c^2}}}.
\label{EQu4.05}
\end{equation}
The square of the invariant spatial interval $\Delta s$ is \cite{BIFinn}
\begin{equation}
(\Delta s)^2=(\Delta x)^2+(\Delta y)^2+(\Delta z)^2-(c/n)^2(\Delta t)^2.
\label{EQu4.06}
\end{equation}
Multiplying the preceding equation by $-1$ and taking the square root of
the result yields another invariant quantity
\begin{equation}
c\Delta \tau =\frac{c}{n}
\sqrt{(\Delta t)^2-(n/c)^2((\Delta x)^2+(\Delta y)^2+(\Delta z)^2)}
\label{EQu4.07}
\end{equation}
from which we obtain the interval of proper time
\begin{equation}
d\tau =\frac{dt}{\gamma n }.
\label{EQu4.08}
\end{equation}
Taking the derivative of the position four-vector (\ref{EQu4.02}) with
respect to
the proper time, we obtain the four-velocity
\begin{equation}
{\mathbb U}=\frac{d{\mathbb X}}{d\tau}
=\frac{d{\mathbb X}}{dt}\frac{dt}{d\tau}
= \gamma c\left (
1,
\frac{dx}{d(ct/n)},
\frac{dy}{d(ct/n)},
\frac{dz}{d(ct/n)}
\right )
\label{EQu4.09}
\end{equation}
and the four-momentum
\begin{equation}
{\mathbb P}=m_0{\mathbb U}= 
m c \left (
1, \frac{dx}{d(ct/n)},
\frac{dy}{d(ct/n)},
\frac{dz}{d(ct/n)}
\right ),
\label{EQu4.10}
\end{equation}
where $m=\gamma m_0$ is the relativistic mass.
The four-force
\begin{equation}
{\mathbb F}=\frac{d{\mathbb P}}{d\tau}
=\frac{d{\mathbb P}}{dt}\frac{dt}{d\tau}
\label{EQu4.11}
\end{equation}
is derived in a similar manner.
\par
In the nonrelativistic limit,
$\gamma=1$, $\tau=t/n$, we find the three-velocity
\begin{equation}
{\bf u}=n{\bf\dot x}=
c\left (\frac{dx}{d(ct/n)},\frac{dy}{d(ct/n)},\frac{dz}{d(ct/n)}\right )
\label{EQu4.12}
\end{equation}
and the three-momentum
\begin{equation}
{\bf p}=nm{\bf\dot x}=
mc\left (
\frac{dx}{d(ct/n)},\frac{dy}{d(ct/n)},\frac{dz}{d(ct/n)}\right )
\label{EQu4.13}
\end{equation}
in a region of reduced light velocity.
\par
For a system of particles, the transformation of the position vector of
the $i^{\rm th}$ particle to $J$ independent generalized coordinates is
\begin{equation}
{\bf x}_i={\bf x}_i(\tau ;q_1,q_2, \ldots, q_J) ,
\label{EQu4.14}
\end{equation}
where $\tau=t/n$.
Applying the chain rule, we obtain the virtual displacement
\begin{equation}
\delta{\bf x}_i=\sum_{j=1}^J
\frac{\partial {\bf x}_i}{\partial q_j}\delta q_j
\label{EQu4.15}
\end{equation}
and the velocity
\begin{equation}
{\bf u}_i=\frac{d{\bf x}_i}{d\tau}=
\sum_{j=1}^J
\frac{\partial {\bf x}_i}{\partial q_j}
\frac{d q_j}{d \tau}
+ \frac{\partial {\bf x}_i}{\partial \tau}
\label{EQu4.16}
\end{equation}
of the $i^{th}$ particle in the new coordinate system.
The relation
\begin{equation}
\frac{\partial{\bf u}_i}{\partial(d q_j/ d \tau)}=
\frac{\partial {\bf x}_i}{\partial q_j}
\label{EQu4.17}
\end{equation}
comes from the derivative of Eq.\ (\ref{EQu4.16}).
Substitution of Eq.\ (\ref{EQu4.17}) into the identity 
\begin{equation}
\frac{d}{d\tau}\left ( m{\bf u}_i\cdot
\frac{\partial{\bf x}_i}{\partial q_j} \right ) =
m\frac{d{\bf u}_i}{d \tau}\cdot
\frac{\partial {\bf x}_i}{\partial q_j}
+
m{\bf u}_i\cdot\frac{d}{d\tau}
\left ( \frac{\partial{\bf x}_i}{\partial q_j}\right )
\label{EQu4.18}
\end{equation}
results in
\begin{equation}
\frac{d{\bf p}_i}{d\tau}\cdot
\frac{\partial{\bf x}_i}{\partial q_j} =
\frac{d}{d\tau}
\left ( \frac{\partial}{\partial(d q_j/d \tau)}
\frac{1}{2}m{\bf u}_i^2
\right ) -
\frac{\partial}{\partial q_j}\left ( \frac{1}{2}m{\bf u}_i^2\right )
\label{EQu4.19}
\end{equation}
by application of the calculus.
\par
For a system of particles in equilibrium, the virtual work of the
applied forces ${\bf f}_i$ vanishes and the virtual work on each
particle vanishes leading to the principle of virtual work
\begin{equation}
\sum_i{\bf f}_i\cdot \delta{\bf x}_i=0
\label{EQu4.20}
\end{equation}
and D'Alembert's principle
\begin{equation}
\sum_i\left ( {\bf f}_i -\frac{d{\bf p}_i}{d\tau}\right )
\cdot \delta{\bf x}_i=0.
\label{EQu4.21}
\end{equation}
Defining the kinetic energy of the $i^{th}$ particle
\begin{equation}
T_i= \frac{1}{2} m{\bf u}_i^2,
\label{EQu4.22}
\end{equation}
we can write D'Alembert's principle (\ref{EQu4.21}) as
\begin{equation}
\sum_j \left [ \left ( \frac{d}{d\tau}
\left ( \frac{\partial T}{\partial(d q_j/d \tau)}\right ) 
-\frac{\partial T}{\partial q_j}
\right ) -Q_j \right ] \delta q_j =0 
\label{EQu4.23}
\end{equation}
using Eqs.\ (\ref{EQu4.15}) and (\ref{EQu4.19}),
where
\begin{equation}
Q_j=\sum_i{\bf f}_i\cdot\frac{\partial{\bf x}_i}{\partial q_j}.
\label{EQu4.24}
\end{equation}
If the generalized forces $Q_j$ come from a generalized scalar potential
function $V$ \cite{BIGoldstein}, then we can write the Lagrange
equations of motion
\begin{equation}
\frac{d}{d\tau}
\left (
\frac{\partial L}{\partial(d q_j/d \tau)}\right ) 
-\frac{\partial L}{\partial q_j} =0,
\label{EQu4.25}
\end{equation}
where $L=T-V$ is the Lagrangian in a linear medium.
\par
For any coordinate $q_j$ that measures a linear displacement of a
particle in a given direction, the partial derivative of the Lagrangian
with respect to that coordinate vanishes.
In that case, substitution of 
\begin{equation}
\frac{\partial L}{\partial q_j}=0
\label{EQu4.26}
\end{equation}
into Eq.\ (\ref{EQu4.25}) produces
\begin{equation}
\frac{d}{d\tau}
\frac{\partial L}{\partial(d q_j/d \tau)}=0.
\label{EQu4.27}
\end{equation}
If the potential energy is velocity independent then
we can define the canonical momentum
\begin{equation}
p_j=\frac{\partial L}{\partial(d q_j/d \tau)}
=\frac{1}{c}\frac{\partial L}{\partial(d q_j/d (ct/n))}
\label{EQu4.28}
\end{equation}
based on the proper temporal integration of Eq.\ (\ref{EQu4.27}).
\par
The consequence that an effective speed of light has for the canonical
momentum is easily illustrated in the case of a nonrelativistic
free particle.
The kinetic energy of the particle, initially at rest in the local
frame, is
\begin{equation}
T=\frac{1}{2}m {\bf u}\cdot{\bf u}.
\label{EQu4.29}
\end{equation}
Then the Lagrangian for the free particle can be written as
\begin{equation}
L=\frac{m}{2}\sum_{j=1}^3 \left ( \frac{dx_j}{d\tau}\right )^2
\label{EQu4.30}
\end{equation}
in rectangular coordinates.
Applying the definition of the canonical momentum (\ref{EQu4.28}), one
obtains
\begin{equation}
p_j=\frac{\partial L}{\partial(d x_j/d\tau)}=m\frac{d x_j}{d\tau}
=mn\frac{d x_j}{dt}
\label{EQu4.31}
\end{equation}
in agreement with the linear momentum of a particle in a linear
medium, Eq.\ (\ref{EQu4.13}).
Although macroscopic particles cannot travel unimpeded through a
material in the continuum limit, light does at reduced speed.
In the next section, the modified Lagrangian dynamics are applied to 
the derivation of macroscopic quantum electrodynamic principles.
\par
\section{Macroscopic Quantization}
\par
Lagrangian dynamics is the basis for the procedure that was used in 
Sec.\ II to quantize the electromagnetic field in a linear medium.
The macroscopically quantized field, Eq.\ (\ref{EQu2.13}), was found to 
be inconsistent with the classical electromagnetic boundary conditions.
In the preceding section, we derived a modification of Lagrangian
dynamics in the context of a uniform $c/n$ speed of light.
Here we show that the quantized field that is derived using the modified
dynamical theory is consistent with the electromagnetic boundary
conditions.
\par
The macroscopic quantization procedure is based on an expansion of the
vector potential in terms of modes as
\begin{equation}
{\bf A}= c\sum_{l\lambda}
q_{l\lambda}(t){\bf u}_{l\lambda}({\bf r}).
\label{EQu5.01}
\end{equation}
Substituting Eq.\ (\ref{EQu5.01}) into the Lagrangian
\begin{equation}
L=\frac{1}{2}\int
\left ( \frac{n^2}{c^2} \left (
\frac{d{{\bf A}}}{d t}\right )^2
-(\nabla\times {\bf A})^2 \right ) d^3{\bf r}
\label{EQu5.02}
\end{equation}
and the wave equation, one obtains the Lagrangian
\begin{equation}
L=\frac{1}{2}
\sum_{l\lambda}
\left (n^2 \dot q_{l\lambda}^2
-n^2\omega_l^2 q_{l\lambda}^2\right ),
\label{EQu5.03}
\end{equation}
as in Section II.
The point of departure for this derivation is the use of
Eq.\ (\ref{EQu4.28}), rather that Eq.\ (\ref{EQu2.05}),
for the conjugate momenta.
Instead of $p_{l\lambda}= n^2\dot q_{l\lambda}$, which leads to 
violation of quantum--classical correspondence, we obtain
\begin{equation}
p_{l\lambda}= \frac{\partial L}
{\partial \left (d q_{l\lambda}/d\tau \right )} =n\dot q_{l\lambda}
\label{EQu5.04}
\end{equation}
for the conjugate momenta.
The effective Hamiltonian
\begin{equation}
H=\sum_{l\lambda}p_{l\lambda} \frac{dq_{l\lambda}}{d\tau}-L
=\frac{1}{2}\sum_{l\lambda} \left ({ p}_{l\lambda}^2+
n^2\omega_l^2 q_{l\lambda}^2\right )
\label{EQu5.05}
\end{equation}
is quantized in the usual way by taking
$P_{l\lambda}=\sqrt{n}p_{l\lambda}$
and $Q_{l\lambda}=q_{l\lambda}/\sqrt{n}$
to be operators satisfying the material-independent commutation 
relations
\begin{equation}
[q_{l\lambda},p_{l^{\prime}\lambda^{\prime}}]=
[Q_{l\lambda},P_{l^{\prime}\lambda^{\prime}}]=
i\hbar\delta_{ll^{\prime}}\delta_{\lambda\lambda^{\prime}}.
\label{EQu5.06}
\end{equation}
Defining the usual material-independent annihilation and creation
operators,
\begin{subequations}
\label{EQu5.07}
\begin{equation}
 a_{l\lambda}=\frac{1}{\sqrt{2\hbar\omega_l}}
(\omega_l Q_{l\lambda}+i P_{l\lambda})
\label{EQu5.07a}
\end{equation}
\begin{equation}
 a_{l\lambda}^{\dagger}=\frac{1}{\sqrt{2\hbar\omega_l}}
(\omega_l Q_{l\lambda}-i P_{l\lambda})
\label{EQu5.07b}
\end{equation}
\end{subequations}
yields the effective Hamiltonian
\begin{equation}
H= \frac{1}{2}\sum_{l\lambda}n\hbar\omega_l
\left ( a_{l\lambda}^{\dagger} a_{l\lambda}+
 a_{l\lambda} a_{l\lambda}^{\dagger}\right ).
\label{EQu5.08}
\end{equation}
In the plane-wave cw limit, the classical energy density is
\begin{equation}
{\cal H}=\frac{1}{2}
\left ( \left ( \frac{n}{c} \frac{d{{\bf A}}}{d t}
\right )^2+(\nabla\times {\bf A})^2 \right )
=\frac{n^2 \omega^2}{c^2}A^2.
\label{EQu5.09}
\end{equation}
Using the classical boundary conditions $A_t=A_v/\sqrt{n}$, the energy
density inside the material is a factor of $n$ greater than the energy
density in the vacuum.
The effective Hamiltonian (\ref{EQu5.08}) exhibits
the conforming enhancement of energy density in a medium.
In contrast, the effective Hamiltonian (\ref{EQu2.12}) in the material,
derived by the original Ginzburg quantization procedure with the boson 
commutation relations, fails the test of quantum--classical
correspondence.
\par
The traveling wave representation of the macroscopic vector potential 
\begin{equation}
 {\bf A}=c
\sum_{l\lambda}
\sqrt{\frac{\hbar}{2n\omega_l V}}
\left ({ a_{l\lambda}}e^{i{\bf k}_l\cdot{\bf r}}+
{ a_{l\lambda}^{\dagger}}
e^{-i{\bf k}_l\cdot{\bf r}} \right )
\hat {\bf e}_{{\bf k}_{l\lambda}}
\label{EQu5.10}
\end{equation}
can be constructed from the relations (\ref{EQu5.07}) between the
canonical variables and the creation and annihilation operators
using the canonical transformation.
The field (\ref{EQu5.10}), like the phenomenological field
(\ref{EQm3.07}), satisfies quantum--classical correspondence with
the electromagnetic boundary condition $A_t=A_v/\sqrt{n}$.
\par
\section{Summary}
\par
The quantized vector potential and the effective Hamiltonian for an
electromagnetic field in a dielectric were derived from the Lagrangian
by the conventional quantization procedure.
This representation of the macroscopically quantized field was shown to
violate quantum--classical correspondence.
The principles of Lagrangian dynamics, including D'Alembert's principle
and the Lagrange equations of motion were re-derived using the
thesis of continuum electrodynamics of a structureless medium that
interacts with light through a constant.
The resulting changes to the principles dynamics in the continuum,
applied in the Ginzburg macroscopic quantization procedure, were found
to be sufficient to repair the violation of quantum--classical
correspondence for macroscopically quantized fields.
\par

\end{document}